\documentclass[journal,transmag]{IEEEtran}
\usepackage[utf8]{inputenc}
\usepackage[TS1,T1]{fontenc}
\usepackage{graphicx}
\usepackage{amssymb}
\usepackage{old-arrows}
\usepackage{cite}
\usepackage{amsmath}
\usepackage{url}

\usepackage{jabbrv}

\usepackage{multirow}
\usepackage{siunitx}

\newcommand{\abs}[1]{\ensuremath{\left \lvert {#1} \right \rvert}}

\newcommand{\vref}{\widetilde{v}}

\DeclareMathOperator{\sgn}{sgn}

\graphicspath{{figs/}}

\begin{document}

\title{Near-Field Millimeter Wave Vector Measurements - Experimental Design \& Measurement Interpretation}

\author{\IEEEauthorblockN{Laurent Chusseau\IEEEauthorrefmark{1}, Thibaut Auriac\IEEEauthorrefmark{1}, Jérémy Raoult\IEEEauthorrefmark{1}}
\IEEEauthorblockA{\IEEEauthorrefmark{1}IES, Université de Montpellier, CNRS, Montpellier, France}}


\maketitle

\begin{abstract}
Near-field imaging experiments exist both in optics and microwaves with often different methods and theoretical supports. For millimeter waves or THz waves, techniques from both fields can be merged to identify materials at the micron scale on the surface or in near-surface volumes. The principle of such near-field vector imaging at the frequency of 60 GHz is discussed in detail here. We develop techniques for extracting vector voltages and methods for extracting the normalized near-field vector reflection on the sample. In particular, the subharmonic IQ mixer imbalance, which produced corrupted outputs either due to amplitude or phase differences, must be taken into account and compensated for to avoid any systematic errors. We provide a method to fully characterize these imperfections and to isolate the only contribution of the near-field interaction between the probe and the sample. The effects of the mechanical modulation waveform and harmonic rank used for signal acquisition are also discussed.
\end{abstract}


%


\section{Introduction}

Near-field imaging is now an essential technique for the characterization of devices and materials. Depending on the targeted application, reflection or transmission modes are used in the infrared \cite{Chen:2019ac}, microwave \cite{Imtiaz:2014aa} or THz \cite{Adam:2011aa}, and for the last two take advantage of the transparency of many materials for non-destructive testing through the surface \cite{Kharkovsky:2007,Plassard:2011,Haddadi:2011,Moon:2014}. 

If high spatial resolution is desired in respect to the measurement wavelength, a probe is required to concentrate the field and overcome the Abbe diffraction limit by means of evanescent waves. At GHz frequencies, the scanning microwave microscope (SMM) frequently uses coaxial resonant probes whose quality factor and resonance can be related to the interaction with the sample \cite{Gao:1998ab,Imtiaz:2012,Monti:2017aa}, or AFM tips \cite{Huber:2012aa,Berweger:2015aa,Horibe:2019aa} that can be modeled electromagnetically in 3D \cite{Oladipo:2013,Wei:2016aa,Wu:2017}. The latter technique is also used in the millimeter wave (mmW) region, which is the realm where it is possible to combine optical and microwave techniques, for example the joint use of propagating waves \cite{Dai:2019aa} and guided waves coupled to specific probes including small antennae  capable of concentrating the electric or magnetic field with a very high resolution \cite{Rosner:2002aa,Novotny:2011,Grosjean:2011}.

Although most of near-field experiments in the mmW have focused on intensity measurements \cite{Nozokido:2001aa,Guillet:2010,Haddadi:2011,Omarouayache:2015c,Chusseau:2017aa,Dai:2019aa}, local vector characterization is necessary to distinguish material changes from loss changes. Many such experiments have recently been proposed in scattering-type scanning near-field optical microscopy (s-SNOM) applied to the THz domain \cite{Ribbeck:2008aa,Keilmann:2009aa,Giordano:2018,Liewald:2018aa}, but only the latter seems to be easy to translate to mmW. We do the same transformation on our 60\,GHz test bench \cite{Chusseau:2017aa}  and chose to include a subharmonic IQ mixer to simultaneously detect the two quadratures of the reflected field, but we kept a complete waveguide configuration. This paper details the new experiment and elaborate the required processing of experimental signals in order to determine without systematic error and with reasonable accuracy the vector voltages and reflection coefficients related to the near-field probe-sample interaction. We show how the rapid phase rotation due to the path length in mmW can be used to achieve these values, and how neutralize the IQ imbalance of the mixer, either in amplitude or phase, by its  theoretical account in data processing. In the end, this gives us an overall figure of merit for the subharmonic IQ mixer and the entire detection system.

The paper is organized as follows. In \S\ref{setup} we describe the experimental setup and its basic operation. \S\ref{process} details the processing to be applied to experimental data, namely the vector voltage determination and how mixer non-ideal behavior transforms the observed vector voltage. We then develop the method for tracing the true reflection coefficient, provided that a quality reference measurement is available. In \S\ref{waveform}, we deal with the effect of the harmonic rank of the detection in relation to the modulation waveform applied to the mechanical modulation of the distance between the probe and the sample, which is a peculiarity of our experiment as compared to s-SNOM. Finally, in \S\ref{IQmix} we qualify the frequency response of our experiment and the subharmonic mixer imbalance.

\section{Measurement Setup}\label{setup}

\begin{figure}
\begin{center}
\includegraphics[width=0.75\columnwidth]{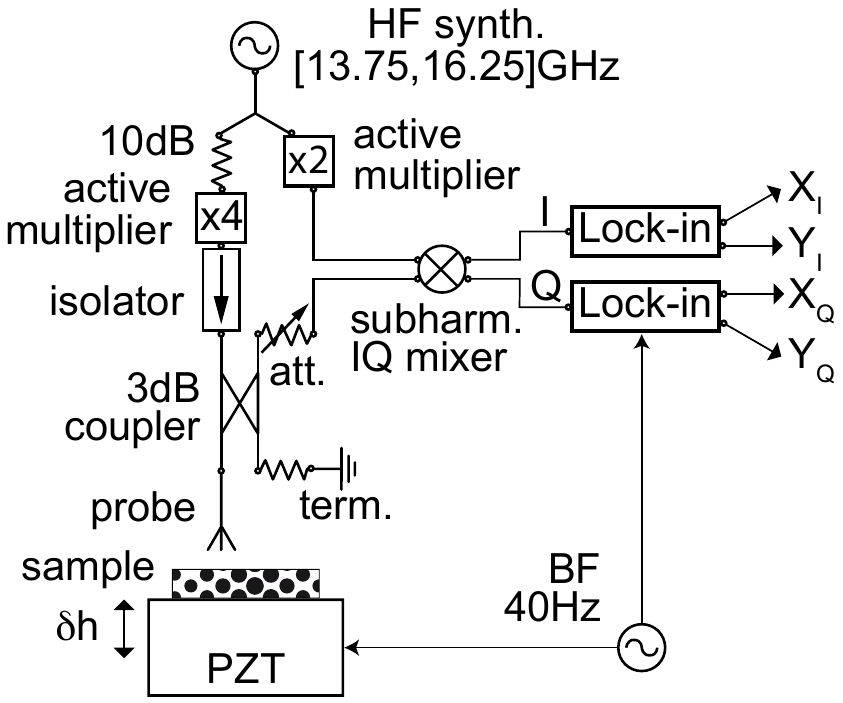}
\caption{Scheme of the mmW near-field experimental setup. All mmW links are made using WR15 waveguides up to the subharmonic IQ mixer.}
\label{experiment}
\end{center}
\end{figure}

The scheme of the new experiment is given in Fig.\,\ref{experiment} with its central element for the vector output: the subharmonic IQ mixer which operates at zero intermediate frequency in the band 55-65\,GHz. Compared to our previous intensity reflectometer \cite{Chusseau:2017aa}, the whole mmW transmitting/detecting system has changed and now uses an HF synthesizer followed by multiplication chains for the detection and LO signal. In the measuring path, an active $\times 4$ multiplier delivering 10\,dBm is used in the front of a 3\,dB coupler which returns the reflected signal at the input of the mixer. 
In between, we take care of the mmW signal level entering the subharmonic IQ mixer by incorporating a variable attenuator to validate the -10\,dBm requirement for linear operation. The LO path of the mixer is fed by an active doubler delivering 10\,dBm. The  $\times 4$ and $\times 2$ active multipliers are not really perfectly flat in frequency, unlike the reference synthesizer. As a result, the subharmonic IQ mixer's conversion loss variations, which are typically 15 to 18\,dB, may experience additional ripple due to these power variations of the mmW sources. The whole system was provided by LTEQ Microwave according to our specifications.

\begin{figure}
\begin{center}
\includegraphics[width=0.7\columnwidth]{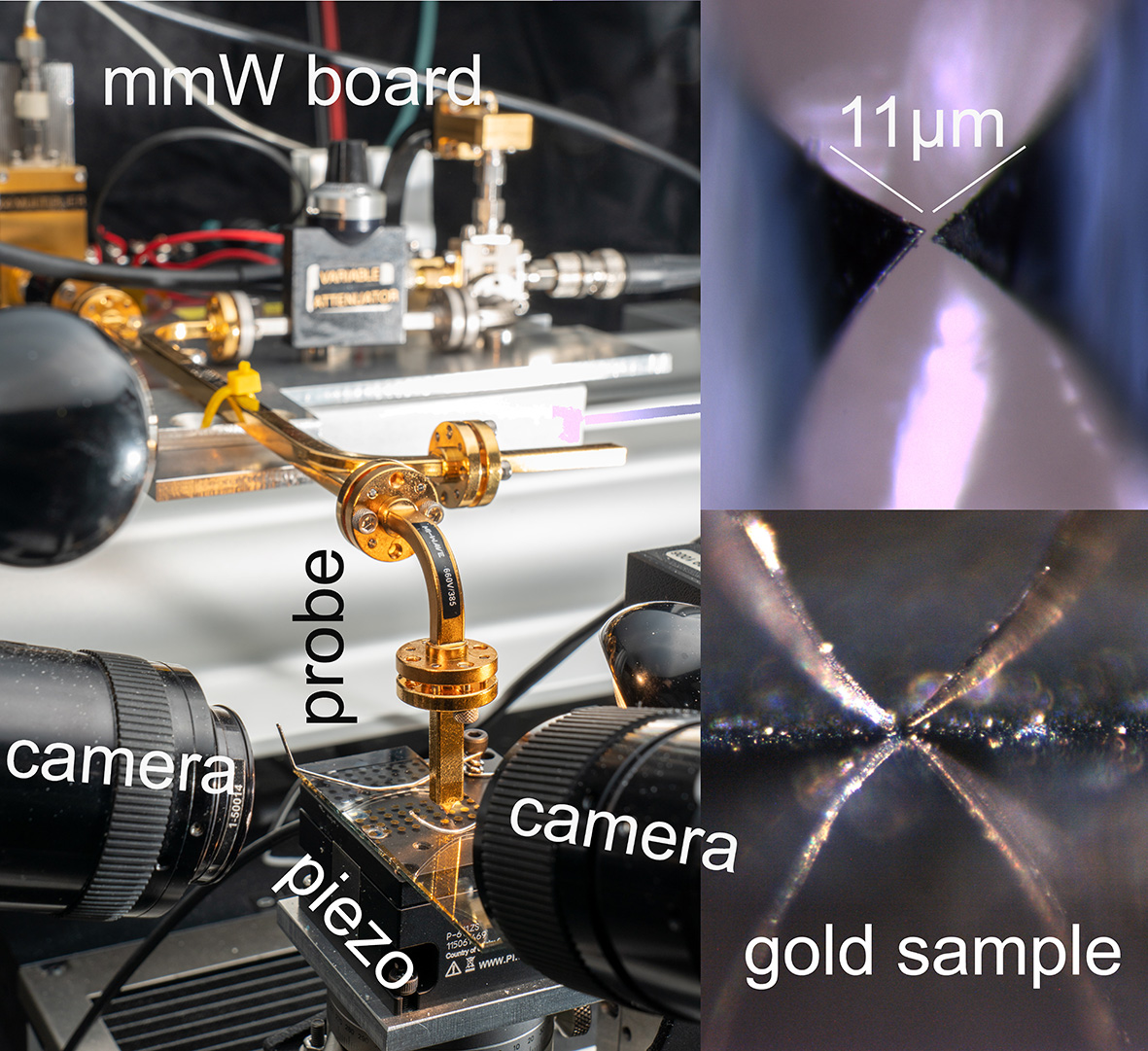}
\caption{Photos of the experiment. Left: emission/detection board in place with the waveguide probe and the two cameras. Right up: tungsten bow-tie probe view from above with the microscope. Right down: Probe as observed by the camera 10\,µm above the gold mirror, the reflection is used to set the minimum distance to $h=10$\,µm.}
\label{photos}
\end{center}
\end{figure}

As in \cite{Chusseau:2017aa}, probes are homemade bow-ties attached to a WR15 open end. Metal triangles used for the bow-tie were produced using fs-laser cut on a 10\,µm thick tungsten sheet. The Fig.\,\ref{photos} illustrates this realization and the probe positioning 10\,µm above a surface. The lateral and vertical positioning over the sample  is ensured by a motorized step-by-step system (Newport XPS-C8) and is optically monitored with two cameras at right angle as shown in the picture. When set at their highest magnification, cameras equipped with Navitar UltraZoom lenses can observe a displacement of 1\,µm, which corresponds to one motor step.

The modulation of the probe-to-sample distance is carried out by the piezoelectric actuator (PI P-611.Z Precision Z-stage) driven by a signal generator whose frequency is set to 40\,Hz due to the limited weight capacity of this actuator. Subharmonic IQ mixer outputs are thus modulated at this low-frequency that is also the reference of the two lock-in (AMETEK 7265 Dual Phase DSP). This allows the filtering of the near-field spatial component ideally split in its real and imaginary parts. In practice both lock-in outputs are complex and refer to the phase of the low-frequency modulation. Correct usage thus requires to set the same phase reference on the two lock-in.

At first glance, the real part in channel I (respectively, the imaginary part in channel Q) of the complex near-field reflected voltage $\vref$ is expected to have the module $m_I=\sqrt{X_I^2+Y_I^2}$ (respectively, $m_Q=\sqrt{X_Q^2+Y_Q^2}$). If these values are obtained straightforwardly, their respective sign must be determined to fully locate $\vref$ in the entire complex plane. 

Our first test sample was chosen to obtain the highest possible near-field reflectivity. It is a gold optical mirror from Edmund Optics with an ultra-flat substrate ($\lambda/20$) and a thin (a few nm) dielectric protective layer. A far-field reflectivity of $\approx 97\%$ is certified in the mid-infrared, a value also expected in the far infrared and in the mmW range since gold is homogeneous and well known \cite{Palik:1998aa}. This sample is intended to be a reference for reflectivity, to which any other sample can be compared.

\begin{figure}[b]
\begin{center}
\includegraphics[width=0.9\columnwidth]{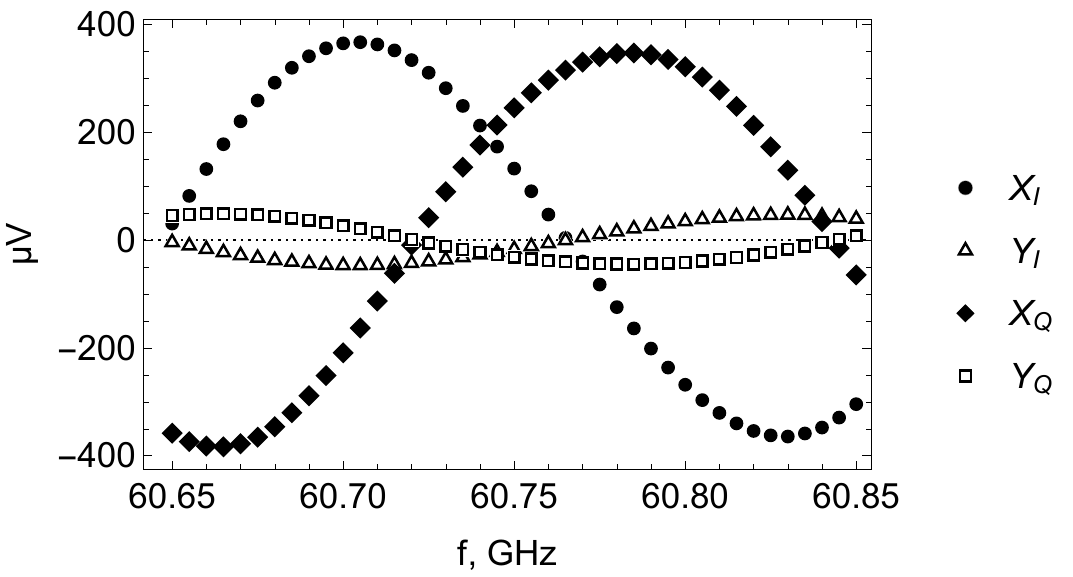}
\caption{Lock-in outputs as measured $h=10$\,µm above a gold mirror with a sinusoidal vibration amplitude $\delta h=20$\,µm while scanning the input frequency.}
\label{xy}
\end{center}
\end{figure}

When scanning the input frequency in a tiny range of 200\,MHz, the four outputs of the lock-ins oscillate rapidly, as shown in Fig.\,\ref{xy}. This is due to the total electrical length of the system here estimated at about $1.25$\,m using the $240 \pm 2$\,MHz period deduced for the $X_i$ and the $Y_i$. 
Since the probe senses the reflection on a flat and homogeneous material, it is not anticipated that it will contribute to a rapidly changing $\vref$ with frequency, but rather that it can be regarded as constant. Consequently, we assume in what follows that the almost sinusoidal evolution of the $X_i$ and $Y_i$ is solely attributed to the electrical length. 

In addition, a close inspection of Fig.\,\ref{xy} shows a deviation from the expected ideal subharmonic IQ mixer: at 60.765\,GHz channel I cancels ($X_I=Y_I=0$) but channel Q is not maximum, which occurs at 60.785\,GHz. Conversely, the same occurs at 60.720\,GHz for channel Q and 60.705\,GHz for channel I. This is quite unexpected for a perfect IQ mixer and shows the mixture of channel signals because of its imperfection. As a result, the measurement of near-field reflection at a single frequency would inevitably be flawed by a systematic error regarding the contribution of the material placed just under the probe. We therefore prefer to carry out frequency sweeps like the one shown in Fig.\,\ref{xy} and process the data to eliminate these errors by assuming that the near-field contribution is constant over the low measurement frequency band.

\section{Measurement Data Processing}\label{process}

\subsection{Vector Voltage Determination}\label{voltage}


Proceeding from a measurement like the one shown in Fig.\,\ref{xy}, the modulus of the reflected near-field voltage is directly obtained by
\begin{equation}
\abs{\vref}=\sqrt{m_I^2+m_Q^2} .
\end{equation}
Accessing the phase is more tedious since the rough estimate $\angle \vref =\arctan \frac{m_Q}{m_I}$ is inherently limited to $[0,\frac{\pi}{2}]$ because of the positive $\arctan$ argument. Going further is possible using the original lock-in outputs, which include the sign changes for channels I and Q when $m_I$ and $m_Q$ go through zero. Noting that these outputs always present opposite $X$ and $Y$ values, the sign are straightforwardly obtained from $\sgn(I) = \sgn(X_I-Y_I)$ and $\sgn(Q) = \sgn(X_Q-Y_Q)$. Keeping track of all these sign changes, we define the phase of the reflected voltage using the heuristic formula
\begin{equation}
\angle \vref = \arctan \frac{\sgn(Q) m_Q}{\sgn(I) m_I} .
\end{equation}

\begin{figure}
\begin{center}
\includegraphics[width=1\columnwidth]{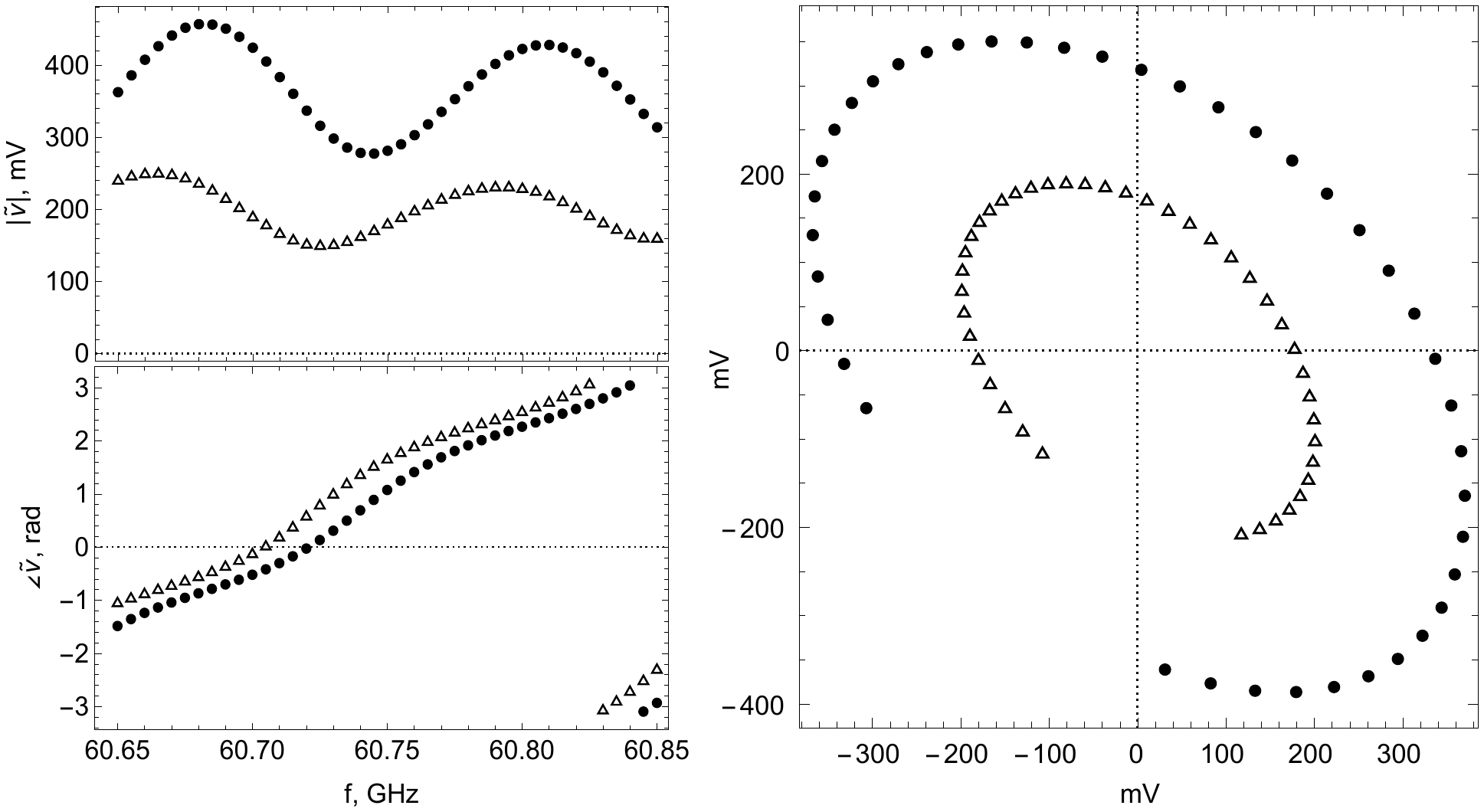}
\caption{Evolutions of the complex reflected near-field voltages $\vref$ for a gold mirror and a GaAs undoped wafer. Left up: modulus, left down: phase, right: complex representation. Dots are for gold and open triangles for GaAs.}
\label{GaAsvsAu}
\end{center}
\end{figure}

We have applied these formulas to the experimental case of the gold mirror in Fig.\,\ref{xy} and also for an unintentionally doped GaAs wafer (ITME, Warsaw). The same geometrical conditions were applied for the measurement, \emph{i.e.} $h=10$\,µm, $\delta h=20$\,µm and a sinusoidal modulation. The results are shown in Fig.\,\ref{GaAsvsAu} with the modulus and phase of $\vref$ and their locations in the complex plane. As expected, $\angle \vref$ covers the whole range $[-\pi,\pi]$ which leads to elliptic representations in the complex plane. 

Although the two materials can be easily distinguished on these curves, especially by their average values $\abs{\vref}$ which are clearly different and by their shifted $\angle \vref$ evolutions, it is impossible at this stage to characterize a material with a single complex value whereas this was hoped by assuming a constant influence of the material on the limited frequency range considered. In this framework, centered ellipses are distorted views of circles usually obtained in microwave by measurements using sliding loads. We attribute this distortion to the imbalance of the IQ mixer and model it in the sequel to refine the analysis.

\subsection{Normalized Vector Reflection}\label{mixer}


Let us consider the ideally reflected complex voltage $v$ in the near-field of the tested surface. As said it should be constant over the tiny frequency range considered because our materials are far from any resonance. The voltage $v$ thus undergoes only a phase shift $\exp j \phi$ due to the electrical length before being transformed by the subharmonic IQ mixer to produce $\vref$. We assume that mixer imbalance is twofold: first, it affects the gain by a factor $\gamma$ possibly different from unity that we apply only to the imaginary part, and second, the imperfect quadrature is accounted for by an additional angle $\theta$ added to the $\pi/2$ phase shift of the Q channel. This is written
\begin{equation}
\vref = \Re(v \exp j \phi) + j \gamma \Im(v \exp j(\phi+\theta)) .
\end{equation}
If we now divide the vector voltage measured on the sample $\vref_S$, for instance the GaAs wafer of Fig.\,\ref{GaAsvsAu} by the vector voltage measured on a reference $\vref_R$, namely the gold mirror in the same figure, which is highly conductive at mmW and supposed to produce the highest possible reflection, we obtain
\begin{equation}\label{cireq}
\frac{\vref_S}{\vref_R}=\rho \frac{\cos(\phi + \varphi) + j \gamma \sin(\phi + \varphi + \theta)}{\cos \phi + j \gamma \sin( \phi + \theta)},
\end{equation}
where $\rho=\abs{v_S}/\abs{v_R}$ and $\varphi = \angle v_S - \angle v_R$. As in reflectometry, the normalized vector reflection $\rho \exp j \varphi$ is thus intended to be the near-field mmW reflection coefficient $\Gamma$ of the tested material, provided that the reference is a perfect metal.

Eq.\,\ref{cireq} is easily identified as a circle in the complex plane parametrized by $\phi$. It is completely described if $\phi \in [0,\pi]$ and collapses to a single point with a perfect IQ mixer ($\gamma=1$ and $\theta=0$). Solving a set of 3 equations for 3 different points such as $\phi=0$, $\phi=\pi/2$ and $\phi=-\theta$ which must belong to the same circle $(X-X_c)^2+(Y-Y_c)^2-R^2=0$, provides the link with our parameters in Eq.\,\eqref{cireq}. After some lengthly developments, we end up with the definition of the circle
\begin{subequations}\label{cirpar}
\begin{align}
X_c&=\rho \cos \varphi\\
Y_c&=\frac{\rho (1 + \gamma^2) \sin \varphi}{2 \gamma \cos \theta} \\
R&=\zeta Y_c 
\end{align}
\end{subequations}
with 
\begin{equation}\label{ideality}
\zeta = \sqrt{ 1 - \frac{4 \gamma^2 \cos^{2}\theta}{\left(1+\gamma^2\right)^2}}.
\end{equation}

\begin{figure}
\begin{center}
\includegraphics[width=0.85\columnwidth]{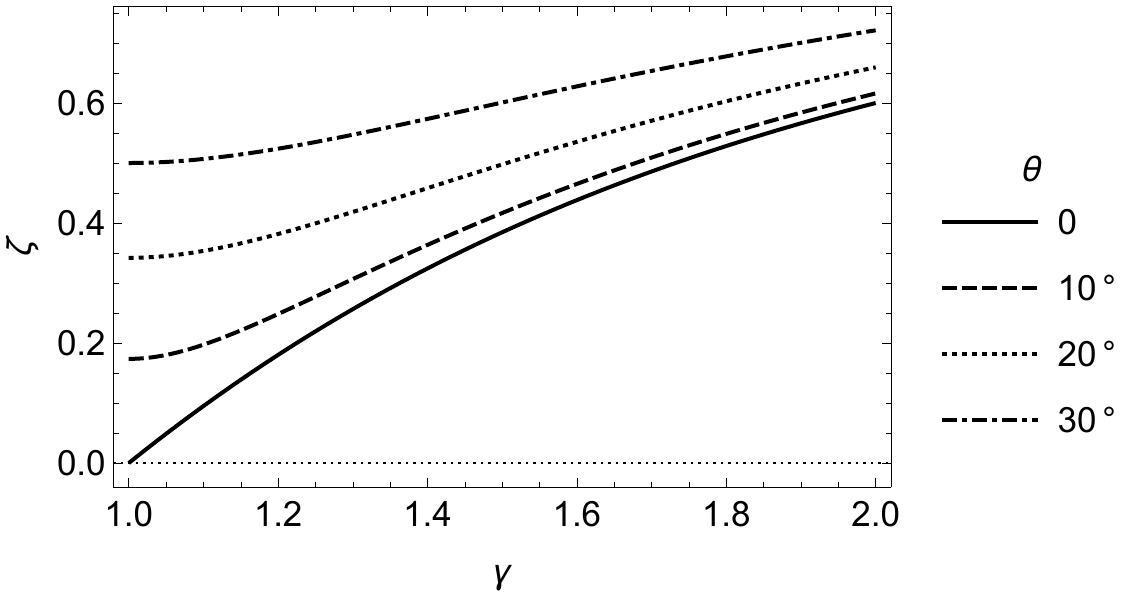}
\caption{Ideality factor $\zeta$ as a function of gain mismatch $\gamma$ and quadrature phase mismatch $\theta$.}
\label{ideal}
\end{center}
\end{figure}

Interestingly the term $\zeta$ is completely independent of both the sample and reference. It is thus some kind of measure of the IQ mixer ideality within our experiment. The gain imbalance $\gamma$ and the quadrature error $\theta$ are mixed in $\zeta$ and cannot be solved separately because we are missing a fourth variable.  Order of magnitude of $\zeta$-values are given in Fig.\,\ref{ideal} as a function of $\gamma$ and $\theta$. For gain errors lower than 3\,dB and phase errors lower than 30° it does not exceed 0.6 which is nonetheless significant compared to its limit value of 1 obtained when $\gamma$ or $1/\gamma \to \infty$ or $\theta \to \pm\pi/2$. We must keep in mind this limit value of 1 which cannot be exceeded under any circumstances in our experiments.   It is a criterion of suitability of the measurements and their exploitation that will be implemented in the following. Values on the order of 0.4 and less, however, seem acceptable.

\begin{figure}
\begin{center}
\includegraphics[width=0.625\columnwidth]{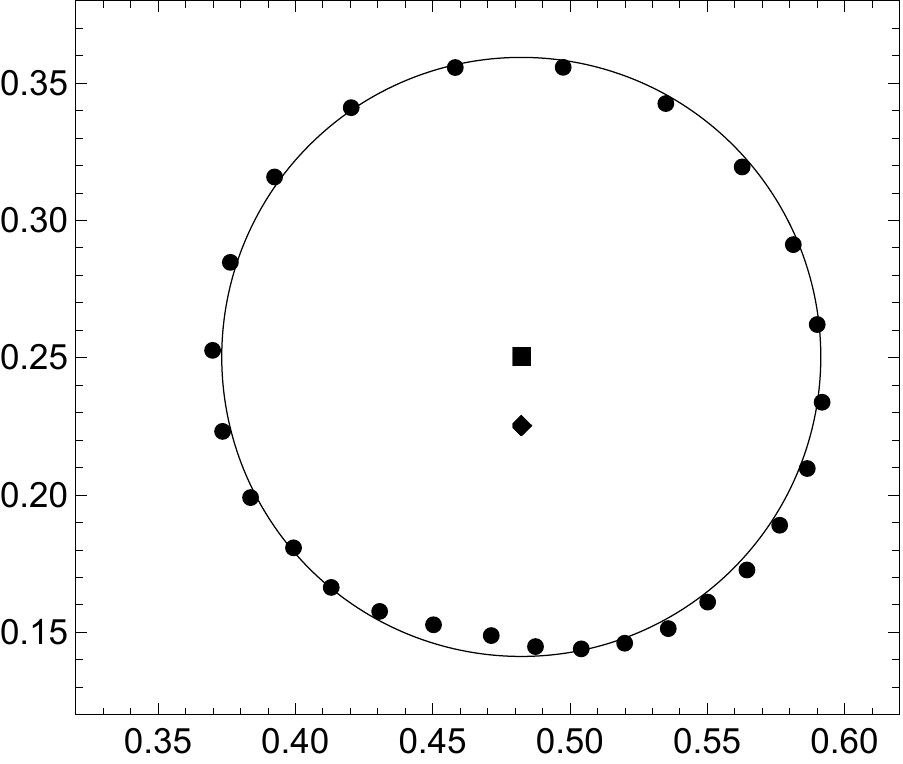}
\caption{Ratio $\vref_S/\vref_R$ in the complex plane for the GaAs sample and the gold mirror reference of Fig.\,\ref{GaAsvsAu}. Dots are measurements, the solid line is the corresponding best circle fit whose center is the filled square, and the filled diamond is the deduced $\Gamma=\rho \exp j \varphi$ normalized vector reflection.}
\label{circle}
\end{center}
\end{figure}

Go back to measurements to calculate and plot the ratio $\vref_S/\vref_R$ from data of Fig.\,\ref{GaAsvsAu}. As shown in Fig.\,\ref{circle} we end up with a nearly perfect circle, just like predicted above. We have limited here the frequency range previously considered to $[60.690-60.815]$\,GHz in order to describe the circle without overlap. Since the points are not evenly distributed on the circle, there is no simple criteria to immediately access the position of its center, so a least-squares fit was used to determine the triplet $(X_c,Y_c,R)$. A secondary result of the fit that minimizes the distance between the measured points and the circle is the error distance accumulated for measurement points to the circle. It is used in the sequel to estimate the uncertainties on each of the inferred values using Student's t-test with 95\% confidence level and classical propagation of errors at the first level on the equations. The error is only $\approx 6\,10^{-4}$ per point in this example, as could be expected given the excellent agreement between the best circle, represented by a solid line in the Fig.\,\ref{circle}, with the measurement points. The position of the center is noted by a filled square.

The module and phase of the near-field reflection are then straightforwardly calculated from Eqs.\,\eqref{cirpar} by
\begin{subequations}\label{final}
\begin{align}
\rho &= \sqrt{X_c^2+Y_c^2-R^2}\\
\varphi &= \arccos \frac{X_c}{\rho} .
\end{align}
\end{subequations}

In the case of Fig.\,\ref{circle} this deduced point is represented by a diamond. It is obviously different from the center, even if according to Eq.\,(\ref{cirpar}a) their real parts are identical. In essence, we end up here with the normalized vectorial reflection $\Gamma=v_S/v_R=(0.482  + j 0.225)\pm 0.004$ with an ideality factor $\zeta=0.436\pm0.005$, which illustrate the effectiveness of the entire measurement operation process, including error estimates. In the ideal case where our reference is perfect, it is expected that $\Gamma$ is the true reflection coefficient in near-field at mmW.

\section{Influence of the mechanical modulation}\label{waveform}

In scattering-type scanning near-field optical microscopy (s-SNOM) the mechanical probe-to-sample distance modulation is always sinusoidal since it is provided by a quartz resonance \cite{Ribbeck:2008aa,Berweger:2015aa,Dai:2019aa,Liewald:2018aa}. 
Unconventional modulation waveforms are possible in our experiment since the mechanical displacement is performed by a piezoelectric actuator controlled by a low-frequency generator. The actuator is optimized for 40\,Hz but has a cut-off frequency of about 100\,Hz, so we can use any arbitrary waveform validating these conditions. Measurements have been performed as test cases with our two samples: the GaAs wafer and the gold mirror.

We try successively sinusoid, square and triangle modulations at constant 40\,Hz fundamental frequency. In each case a measure is done on the reference and reproduce in exactly the same conditions for position, modulation and detection for the GaAs sample. Initial position of the probe is always the minimum distance $h=10$\,µm above the surface before applying the modulation whose amplitude is intended to be $\delta h=20$\,µm, as calibrated for sinusoid. Depending on the modulation waveform, the effective $h$ may be smaller, especially with square modulation where the transfer function of the piezoelectric actuator exhibits an overshoot. Although the displacement approximately follows the command, the rise and fall times are not instantaneous and the effective distance between the probe and the sample is distorted. This occurs also to a lesser extent with triangular modulation.

\begin{table*}
\caption{Deduced normalized reflection coefficient $\Gamma$ of GaAs versus the modulation waveform and the harmonic rank.}
\begin{center}
\begin{tabular}{c|c S|c S|c S}
\multirow{2}{*}{Rank}&\multicolumn{2}{c|}{Sinus Modulation}&\multicolumn{2}{c|}{Square Modulation}&\multicolumn{2}{c}{Triangle Modulation}\\
\cline{2-7}
&$\Gamma$&$\zeta$&$\Gamma$&$\zeta$&$\Gamma$&$\zeta$\\
\hline
1& $(0.482+j 0.225) \pm 0.004$ & 0.436 & $(0.489+j 0.226) \pm 0.004$ & 0.435 & $(0.487+j 0.226) \pm 0.005$ & 0.436 \\
2& $(0.655+j 0.280) \pm 0.006$ & 0.408 & $(0.679+j 0.270) \pm 0.009$ & 0.440 & $(0.687+j 0.283) \pm 0.006$ & 0.410 \\
3& --- & 5.6 & $(0.511+j 0.231) \pm 0.005$ & 0.436 & $(0.528+j 0.234) \pm 0.004$ & 0.431 \\
4& --- & 1.00 & $(0.728+j 0.289) \pm 0.005$ & 0.414 & $(0.735+j 0.283) \pm 0.021$ & 0.418 \\
5& & & --- & 2.9 & --- & 2.9 \\
6& & & $(0.760+j 0.263) \pm 0.017$ & 0.418 & & \\
7& & & $(0.411+j 0.173) \pm 0.010$ & 0.452 & & \\
8& & & $(0.779+j 0.276) \pm 0.026$ & 0.428 & & \\
9& & & $(0.355+j 0.146) \pm 0.021$ & 0.47 & & \\
\hline
\end{tabular}
\end{center}
\label{mytable}
\end{table*}

\begin{figure}[b]
\begin{center}
\includegraphics[width=0.7\columnwidth]{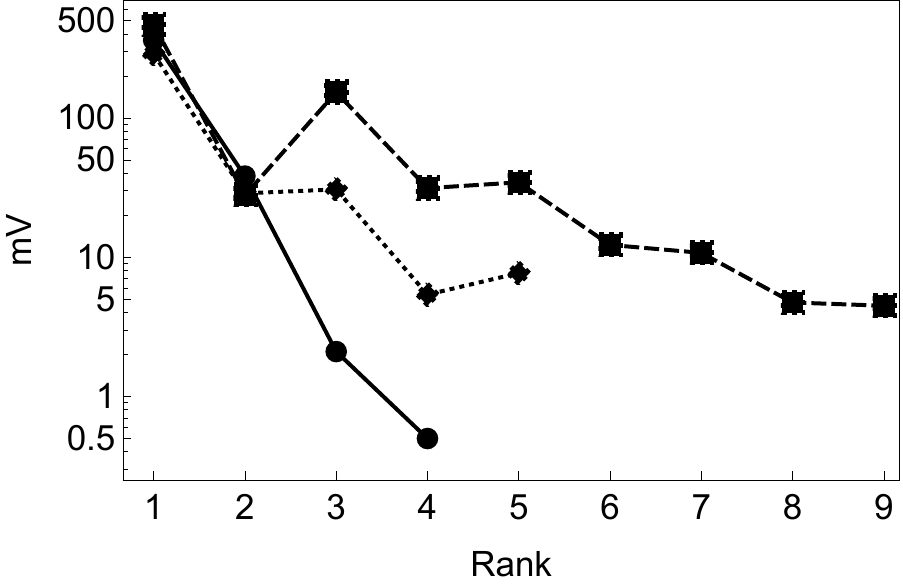}
\caption{Average module of the voltage detected $\abs{\vref}$ for $f \in [60.690,60.815]$\,GHz versus harmonic rank and modulation waveform: {\large$\bullet$} are for sinus, $\blacksquare$ for square and $\blacklozenge$ for triangle modulations.}
\label{hampl}
\end{center}
\end{figure}

The experimental procedure and the determination of the vectorial voltage described in Fig.\,\ref{GaAsvsAu} were reproduced for the three modulations with increasing harmonic ranks selected on the lock-ins. This was done as long as the detected signal emerged from the noise. As can be seen from the average $\abs{\vref}$ voltages plotted in Fig.\,\ref{hampl}, the detected signal decreases rapidly with harmonic rank, but it also depends a lot on the waveform of the modulation. Nevertheless, $\abs{\vref}$ is very similar for all waveforms at the first two harmonics. Beyond that, the decrease in intensity is huge for sinusoidal modulation, which generates those high harmonics only by the nonlinear probe-to-sample interaction caused by the rapid decrease of intensity in the near-field. This peculiarity of probe-to-sample interaction is often used to improve the spatial resolution \cite{Adam:2000,Knoll:2000,Chusseau:2017aa}. For other modulation waveforms, higher harmonics are less attenuated, especially in the case of square modulation. Actually they tend to follow the intensities that could be computed by the Fourier transform of the initial waveforms, in particular the high attenuation of even harmonics. For such waveforms we can no more argue for an improvement in resolution by detecting high order harmonics.

\begin{figure}[b]
\begin{center}
\includegraphics[width=0.85\columnwidth]{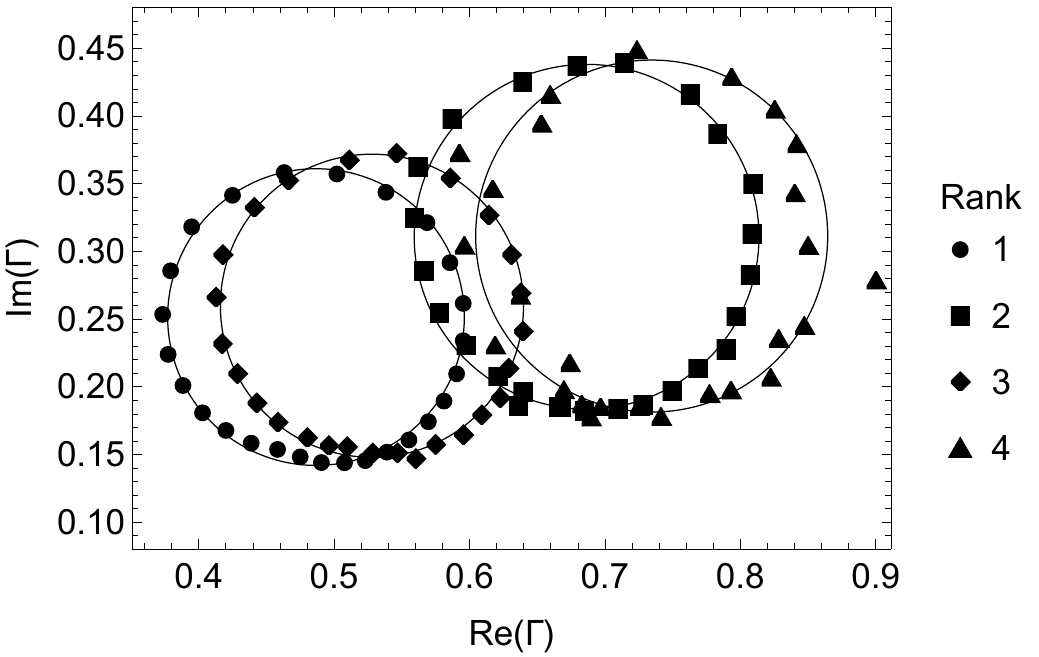}
\caption{Normalized reflection measured for the four first harmonic ranks with a triangle modulation waveform. Dots are measurements and circles their corresponding best fits.}
\label{trcirc}
\end{center}
\end{figure}

The results were interpreted using Eqs.\,\eqref{final} applied to the best fitted circles on the experiments. We used $\zeta$ as a determinant: if its value exceeds 1, which according to the discussion in \S\ref{mixer} is impossible in the framework of our subharmonic IQ mixer model, then we rejected the measure as a whole. The practical reasons for such a mismatch can be multiple, such as non-reproducible positioning over our test surfaces, weak or noisy signals... The factor $\zeta$ is therefore an effective quality criterion for the control of our measurements. In the case of triangle modulation, the Fig.\,\ref{trcirc} illustrates the measurement points with the 4 circles obtained for the first 4 harmonics. Although the measurement points are all well located on circles, it is obvious that the even and odd harmonics are grouped together, but the two groups do not merge even if they overlap.  
As a result the extracted reflection coefficient with even harmonics shows real and imaginary parts which both have higher values.

The complete results for all harmonics of all waveforms are given in the Table\,\ref{mytable}. Rejected cases occurring because $\zeta \ge 1$ are represented by a dash and this occurs especially for all harmonics 5. We believe that this is because this frequency of 200\,Hz is the only one that is also a harmonic of the power supply frequency, which inevitably produces a large increase of noise in the detection circuit. As soon as $\abs{\vref}$ is high enough in all other cases, except for the higher harmonics of the sinusoidal modulation, we obtain $\zeta \approx 0.4$ and the results do not seem to be biased.

\begin{figure}
\begin{center}
\includegraphics[width=\columnwidth]{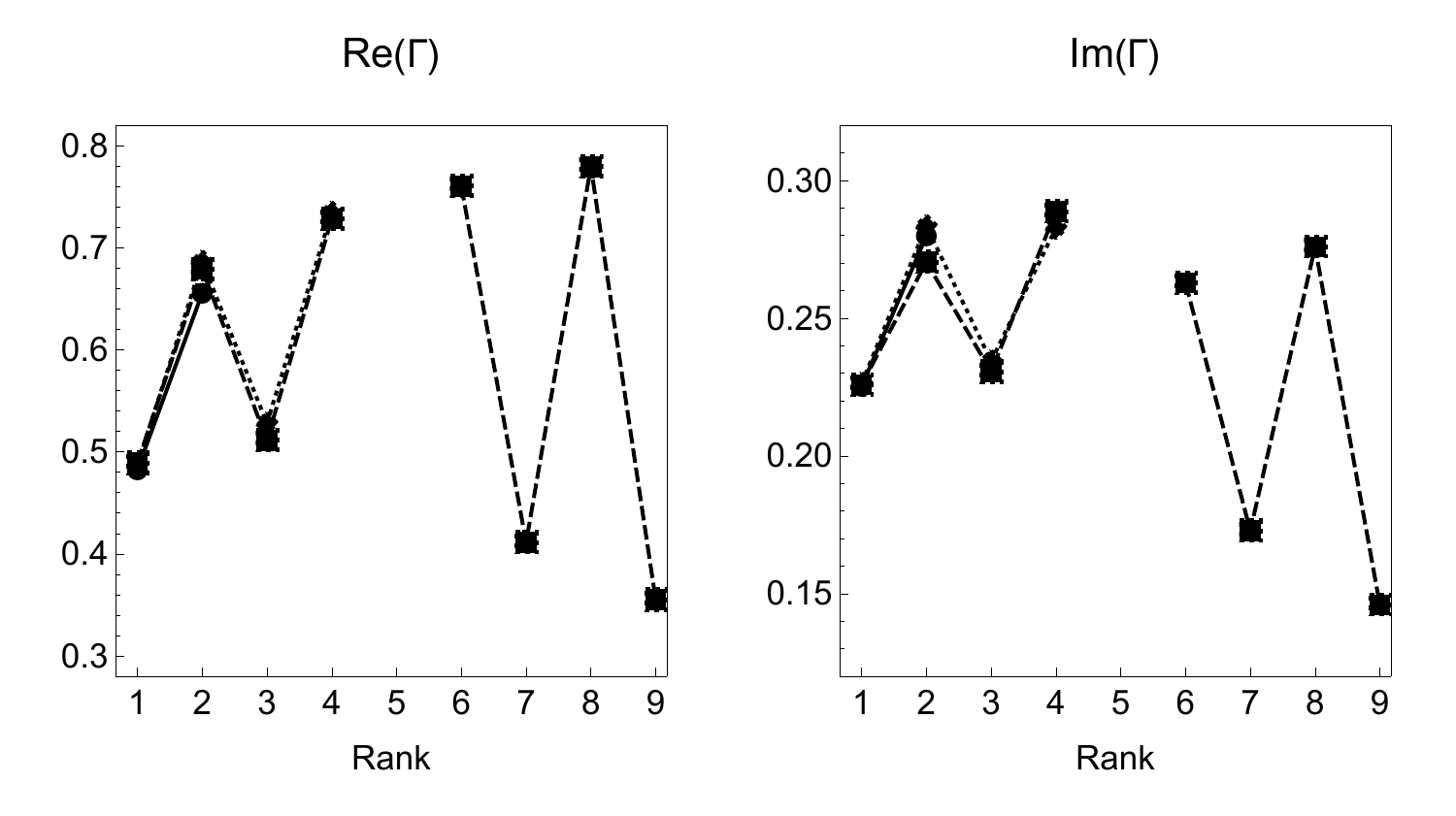}
\caption{Real (left) and imaginary parts (right) of the normalized reflection $\Gamma$ versus harmonic rank and modulation waveform: {\large$\bullet$} are for sinus, $\blacksquare$ for square and $\blacklozenge$ for triangle modulations.}
\label{reimhrank}
\end{center}
\end{figure}

The data of Table\,\ref{mytable} are plotted in Fig.\,\ref{reimhrank}. The superposition of the results for all waveforms highlights the experimental independence in relation to this setting. On the contrary, the alternation of the values between even and odd harmonics is conspicuous.  While the real parts have values that increase with the rank of even harmonics, the opposite occurs for odd harmonics, and the same trend is practically true for the imaginary part. Nevertheless the most notable thing is the dichotomy between even and odd harmonics. The most surprising fact of these results is that for a given harmonic rank the normalized reflection value determined is the same regardless of the modulation shape, and this is verified within the limit of the error bars for the first two harmonics and still very close for next two one. Note that increasing the harmonic rank rapidly decreases the signal and therefore mechanically increases the error.

Obtaining different $\Gamma$ values from one rank to another is deeply counter-intuitive, as is the fact that even and odd harmonics seem to behave identically. Actually, if only the material tested in the near-field matters, the reflection should be expressed from the dielectric constant $\epsilon$ of the material as predicted by the Fresnel coefficients \cite{Novotny2006}, $\Gamma=(\epsilon-1)/(\epsilon+1)$. For instance with sinusoidal modulation at rank 1, it leads to $\epsilon=2.863 + j 1.581$, \emph{i.e.} a refractive index of $1.751 + j 0.451$, which is far from the tabulated value of $3.606 + j 0.0012$ at 300\,GHz \cite{Palik:1998aa}.

Therefore, the quantities determined here must not be regarded as reflection coefficients in the usual sense of microwaves and it is not realistic to extract any $\epsilon$ from them. Although we use the oscillation of the probe-to-sample distance, which concentrates the useful signal on the near-field, we had already shown by approach curves with these bow-tie probes that a longer distance interaction also exists \cite{Chusseau:2017aa}, which certainly  enters here in the final quantitative result. The probable origin of this interaction was identified and attributed to the mandatory openings between the two triangles constituting the bow-tie, and we had also noticed that such a long-distance contribution may be favored if a large dielectric support is used to support the two triangles. For the present probe all precautions had been taken to minimize these effects.  Nonetheless, the quantities extracted are not reflection coefficients even if we used the usual $\Gamma$ notation of microwaves. They seem however characteristic of the materials placed in the near-field of our bow-tie probe and are obtained with a very good accuracy.

\section{Measurement of subharmonic IQ mixer ideality}\label{IQmix}

Since the $\zeta$ factor is a measure of the ideality of our instrumentation, it is worth characterizing it over the entire available bandwidth. Under conditions of sinusoidal modulation with $h=10$\,µm and $\delta h=20$\,µm and detection on the fundamental frequency, we therefore measured both our gold mirror reference and our GaAs sample from 55 to 65\,GHz with a 10\,MHz stepsize. We then proceed to the extraction defined in \S\ref{process} to obtain $\zeta$ and $\Gamma$ using a 120\,MHz sliding window which allows the application of Eqs.\,(\ref{cirpar}-\ref{final}). 

\begin{figure}
\begin{center}
\includegraphics[width=0.7\columnwidth]{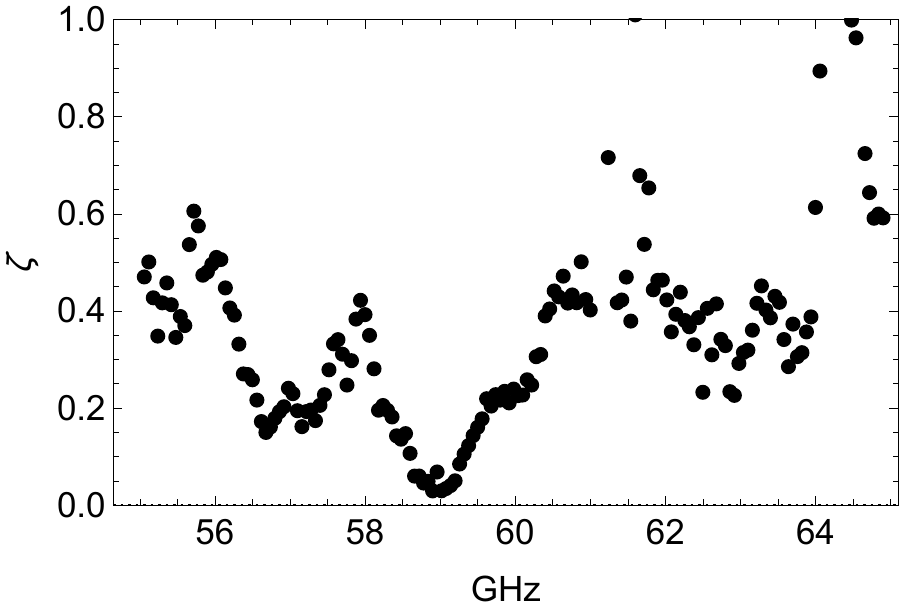}
\caption{Ideality factor $\zeta$ versus frequency.}
\label{zeta}
\end{center}
\end{figure}

Figure\,\ref{zeta} shows the evolution of $\zeta$ as a function of frequency over the entire operating band of the subharmonic IQ mixer. Obviously, $\zeta$ varies a lot in the whole band with really high values above 61\,GHz which exceed unity around 64GHz. This is in full agreement with the observation of a drastic drop of the measured $\vref$ at such frequencies for both gold and GaAs. It is attributed to the conjunction of a power drop of the two multipliers, especially the one dedicated to the LO which provides 3dBm less than below, and an increase in the conversion losses of the subharmonic IQ mixer, leading to an almost $\vref$ cancellation at lock-in outputs. Figure\,\ref{zeta} shows that the best frequency range for our system is around 59\,GHz, but if we assume a maximum value of $\approx 0.4$ it is still quite usable from about 56\,GHz to 61\,GHz. We set out this criterion here because although the frequency around 60.75\,GHz used in \S\ref{process} and \S\ref{waveform} is at the limit of this useful band, we saw that the results were quite satisfactory and accurate.

It should be noted here that due to a handling error, the spacing between the probe tips has been modified since the recordings in \S\ref{process}. If this spacing was previously 11\,µm, it is now 15\,µm due to a too abrupt touch that exceeded the tungsten elasticity limits. Nevertheless the probe has been repaired and kept its global shape with only this new spacing parameter changed. A fundamental consequence of this probe crash and repair is that $\zeta$ is independent of this spacing since we obtain here in Fig.\,\ref{zeta} $\zeta=0.433 \pm 0.024$  at 60.75\,GHz  whereas it was determined in \S\ref{waveform} at $\zeta=0.436 \pm 0.005$ (see Table\,\ref{mytable}). The parameter $\zeta$ is therefore a characteristic that is essentially associated with the mixer imbalance but which nevertheless includes any possible variations in LO power that affect its operating point. 

\begin{figure}
\begin{center}
\includegraphics[width=0.85\columnwidth]{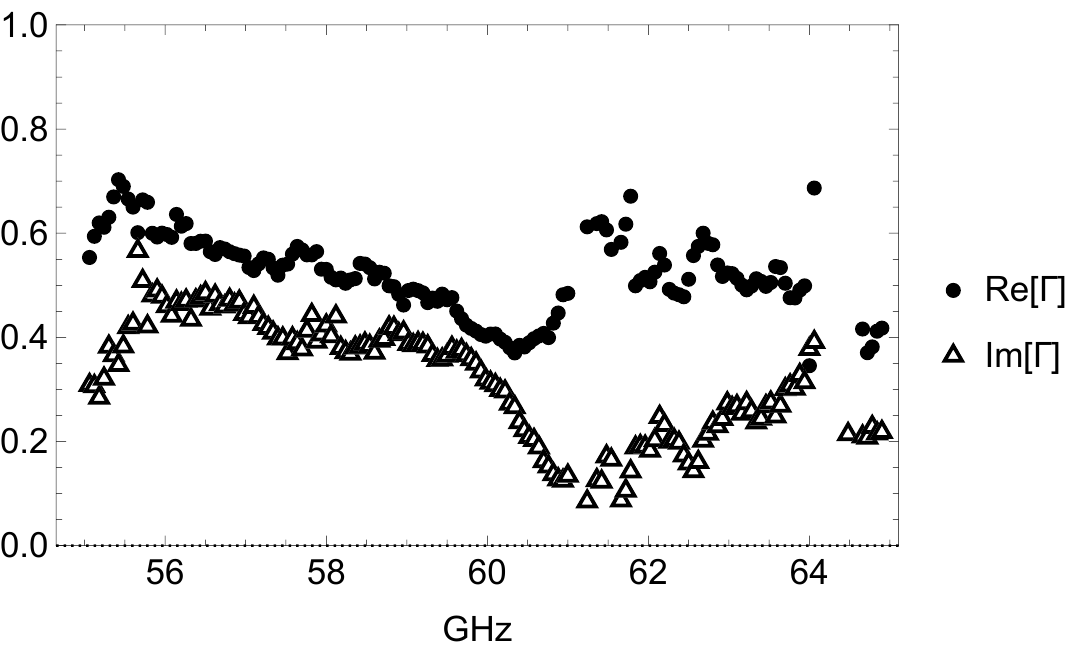}
\caption{Real and imaginary parts of the normalized reflection $\Gamma$ versus frequency.}
\label{reffreq}
\end{center}
\end{figure}

The determination of $\zeta$ over the whole frequency band also yields $\Gamma$ thanks to  Eqs.\,\eqref{final}. It is plotted in Fig.\,\ref{reffreq} where frequencies leading to $\zeta \ge 1$, for instance around 64\,GHz, have been rejected. If $\Gamma$ really represented the reflection coefficient only related to the interaction of a dipole with a homogeneous material, we should expect a  value corresponding to the Fresnel reflection coefficient, $\Gamma=(\epsilon-1)/(\epsilon+1)$, which should be constant since both gold and GaAs are not dispersive at mmW. This is not the case here, even if in the whole range 56-60\,GHz the values vary very little. Since, on the other hand, $\zeta$ encompasses all the imperfections of the experiment involving the subharmonic mixer and its LO, this variation can only be attributed here to the interaction of the probe and the sample which does not seem to follow a simple model. In truth this is confirmed by the fact that we have here at 60.75\,GHz $\Gamma=(0.400 + j 0.152) \pm 0.014$ instead of $\Gamma=(0.482 + j 0.225) \pm 0.004$ found previously before the probe was modified. 

We can therefore conclude that our near-field measurement remains dependent on the probe used, even if the uniqueness of $\zeta$ has shown that all other sources of variability related to the experiment have been compensated for. This is clearly a drawback as we cannot trace back to universal material characteristics but it is probably still possible to distinguish near-field materials after learning with each probe used. In future work, we will try to demonstrate and illustrate this behavior with other probes and materials.

\section{Conclusion}

We presented the realization of a new measurement bench for near-field reflectometry in mmW. A detailed analysis of the operation of this bench has been carried out, making it possible to take advantage of the electrical length that exists between the test and measurement paths. We illustrate its use to extract the near-field contribution that is independent of the mmW board from measurements distorted by systematic errors. A spin-off of this theoretical analysis of the operation of the experiment is the definition of a coefficient of ideality $\zeta$ to which the IQ mixer imbalance and the gain variations of the mmW sources mainly contribute.

We applied this technique to the near-field measurement of a GaAs substrate and a gold mirror. At fixed frequency the study shows a result that is substantially independent of the waveform of the mechanical modulation but not of the order of the harmonic used for detection. Curiously the even and odd harmonics lead to two groups of results that are quite consistent with each other but remain unexplained.

A study of our measurement system  over the entire 55-65\,GHz band has allowed us to qualify $\zeta$. For reasons related to our equipment, the latter one turns out to be rather bad in high frequency but allows very reliable measurements between 56 and 61\,GHz. Moreover, $\zeta$ appears independent of the probe, which is not the case of the normalized vectorial voltage $\Gamma$ measured in the near-field. It is clear that our probe does not have a trivial interaction with the sample and is responsible for this unexpected variation. In the past, we had already observed similar effects with a long range detection effect from the lateral radiation of our probes \cite{Chusseau:2017aa}. However, here we have taken great care to minimize it by eliminating any protruding dielectric material used for mounting the probe. The values obtained for $\Gamma$ and its evolution with frequency cannot be explained with a simple model of the probe-sample interaction. Linking $\Gamma$ to the material parameters of the sample under test in the near-field will therefore require a further study, which can certainly only be carried out through full 3D electromagnetic modeling of the coupled probe-sample system.

Experimentally, the deduced $\Gamma$ values remain reproducible and therefore characteristic of the material for a given probe. In future work, they will be used to distinguish materials in the near field, which probably implies prior calibration and learning procedures. The sub-wavelength spatial resolution that has been demonstrated for intensity measurement should also be studied in this context.


\end{document}